\documentstyle[aps,prl,twocolumn,graphicx]{revtex}

\newcommand{\CGO}{$\rm CuGeO_3\,$}

\def\be{\begin{equation}}
\def\ee{\end{equation}}
\def\bea{\begin{eqnarray}}
\def\eea{\end{eqnarray}}
\def\vecs{{\bf S}}

\begin{document}

\draft

\title{Finite Temperature DMRG Investigation of the 
Spin-Peierls Transition in CuGeO$_3$}

\author{A. Kl\"umper, R. Raupach, F. Sch\"onfeld}
\address{Institut f\"ur Theoretische Physik, Universit\"at zu K\"oln,
Z\"ulpicher Str. 77, 50937 K\"oln, Germany}



\address{~\vspace{0.5cm}\\ \parbox{14cm}{\rm We present a 
numerical study of thermodynamical properties of 
dimerized frustrated Heisenberg chains down to extremely low temperatures 
with applications to \CGO. A variant of the finite temperature density matrix 
renormalization group (DMRG) allows the study of the dimerized phase
previously unaccessible to {\it ab initio} calculations. We investigate
static dimerized systems as well as
the instability of the quantum chain towards lattice dimerization.
The crossover from a quadratic response in the free energy to the distortion field 
at finite temperature to nonanalytic behavior at zero temperature is 
studied quantitatively.
Various physical quantities are derived and compared with experimental data for \CGO 
such as magnetic dimerization, critical temperature, susceptibility and entropy.}}
\maketitle


\section{Introduction}

The investigation of low dimensional quantum spin systems
has attracted widespread and general interest. Active
research is performed on a large class of magnetic materials,
experimentally as well as theoretically. 
Up to date, an enormous understanding of the low-energy physics of 
quasi-one dimensional systems such as spin-Peierls, Haldane 
and ladder systems has been reached. Still it is
a challenging task to investigate especially the low temperature 
regime of these systems, because for $T$ close to zero thermal 
fluctuations are strongly reduced and quantum fluctuations become 
dominant. 

The recently developed transfer-matrix DMRG, 
which combines White's DMRG idea \cite{White92} with the 
transfer-matrix approach \cite{SuzukiI87}
is particularly suited for this task 
\cite{Bursill96,Wang97,Shibata97}, because the thermodynamic limit in
quantum space can be performed exactly.
Most other numerical techniques such 
as the ``traditional'' Hamiltonian DMRG or exact diagonalization are 
restricted to either zero temperature or to rather large temperatures. 
Quantum Monte Carlo algorithms \cite{MonteCarlo} also can only 
be applied to systems
finite in both quantum and Trotter direction. Furthermore, the sign problem
for frustrated systems
is a rather sophisticated point associated with the latter approach.

In this paper we investigate the thermodynamics of dimerized frustrated 
Heisenberg chains, which are believed to model appropriately the spin
degrees of freedom of, for instance, the inorganic spin-Peierls compound \CGO.
The purpose of this paper is twofold. First, we want to show 
that a density matrix 
renormalization group treatment of a ``lattice path integral formulation''
for general spin chains with dimerization and frustration is possible. 
Within this 
approach we will understand qualitatively and quantitatively the instability 
of one-dimensional frustrated antiferromagnetic Heisenberg systems towards 
dimerization at finite temperature. Second, we want to show that
many of the thermodynamic properties of \CGO can be understood 
within the one-dimensional adiabatic approach.
At least with respect to the location of $T_{\mbox{\tiny SP}}$, 
the entropy and the susceptibility down to temperatures of 5K the role 
of higher dimensional magnetic couplings is less relevant.

The paper is organized as follows. 
Computational aspects and the model Hamiltonian are given in section II.
In section III we analyze the 
dependence of the free energy on the dimerization 
and the crossover from analytical to non-analytical behavior 
in the limit of 
vanishing temperature. We compare our numerical findings to 
theoretical results based on scaling arguments.
In addition we present the temperature dependence of the order parameter 
within the adiabatic approach.
In section IV we show results for the magnetic
susceptibility as well as the entropy which 
compare excellently with the experimental data for \CGO. 
Summary and conclusion are given in section V.

\section{model and method}
The dominant magnetic interactions in $\rm CuGeO_3$ are due to Heisenberg
spin exchange between $Cu^{2+}$ ions along the $c$-axis of the crystal
with strong nearest ($J$) and next-nearest neighbor ($\alpha J$) 
interactions \cite{Geertsma96}. 
As usual, we assume a linear dependence of the nearest neighbor
exchange integral on the local structural deformation. The commonly 
used Hamiltonian in adiabatic approximation reads
\be
\hat{H}=\sum_i \left\{ J((1+\delta_i)\vecs_i \vecs_{i+1}+
\alpha\vecs_i \vecs_{i+2})
+\frac{K}{2}\delta_i^2 - h S_i^z \right\},\label{Ham}
\ee
where $\vecs_{i}$ are spin 1/2 operators, $\delta_i = (-1)^i \delta$ 
denotes the modulation of the exchange 
coupling in the dimerized phase ($\delta > 0$ is assumed)
and $h=g \mu_B H_{\mbox{\tiny ex}}$ represents the effective external magnetic field. The last two terms in 
(\ref{Ham}) are the elastic energy associated with 
the lattice distortion and the Zeeman energy.

In order to have a translationally invariant description we combine
each two spin 1/2 to an effective four state object resulting in a
model with only nearest
neighbor interaction. A similar strategy has been adopted in Ref. 
\cite{Maisinger98}
to cast a system with next-nearest neighbor interaction in a form
suitable for applications of the transfer matrix DMRG algorithm. 
Firstly, the elastic part of the Hamiltonian is neglected 
pending further notice, i.e. $K=0$, since it only enters as a trivial
constant in all energies and can be added finally.

After chequerboard decomposition \cite{Suzuki76} of the effective four
state problem the
partition function is calculated by means of the quantum transfer 
matrix $T_{M}$ \cite{Suzuki76,Betsu85}. 
In the thermodynamic limit all
physical quantities are determined by the largest eigenvalue $\Lambda$
of $T_{M}$ and the corresponding left and right eigenstate. More precisely, 
the free energy per site of the spin-1/2 chain
reads $f=-\frac{T}{4}\ln\Lambda$.
Expectation values of local operators like
the internal energy $u$, magnetization $m$ or nearest neighbor
correlation $\left<\vecs_{i}\vecs_{i+1}\right>$ etc. can be expressed in terms of the
eigenstates (see \cite{Wang97} and for a general reference to 
transfer matrix formalisms e.g. \cite{ExpVal}). 
The quantum transfer matrix $T_{M}$, $\Lambda$ and the eigenvectors are
calculated by iterative augmentation in Trotter direction within
an application of the DMRG concept as in \cite{Wang97,Shibata97}. For all
calculations we kept 24 states in the renormalization step and chose
$(T M)^{-1}=0.05$.

\section{general results}

First we calculate the free energy $f(T,\delta)$ for the system (\ref{Ham})
for a series of fixed distortions $\delta_i=(-1)^i\delta$.
From general principles we expect analytic behavior for
any finite temperature, i.e.
\be
f(T,\delta)=f(T,0)+\frac{1}{2!}a(T)\delta^2
+\frac{1}{4!}b(T)\delta^4+O(\delta^6),
\label{expans}
\ee
where the odd powers in $\delta$ disappear due to the obvious symmetry 
($\delta \to -\delta$).
The coefficient $a(T)$ of $\delta^2$ is the reciprocal of the static
``dimerization susceptibility''.
Hence, a zero in this coefficient signals a structural transition.

Numerically we determine the coefficient by analyzing 
\be
A(T,\delta)= \frac{1}{J} \left[ -\frac{1}{\delta}\frac{\partial f(T,\delta)}{\partial
  \delta} + K \right]
= \frac{\langle\vecs_1  \vecs_2\rangle - 
        \langle\vecs_2  \vecs_3\rangle}{2\,\delta}
\label{qt}
\ee
which in the limit $\delta \rightarrow 0$ is simply denoted
by $A(T)$. 
Note that $A$ is independent of $K$.
In Fig.1 we show DMRG results for the temperature dependence of $A(T,\delta)$ 
for various $\delta$ in the case $\alpha = 0$.

The relation $a(T)=K-J A(T)$ holds exactly, but
we are restricted to finite $\delta$ in our numerics since the
error in $A(T,\delta)$ scales with $1/\delta$. (Determining $A(T)$
directly from $f(T,\delta)-f(T,0)$ would even involve an error of 
order $1/\delta^{2}$.) In the sense of a multi-point formula
we can assume an expansion like (\ref{expans})
up to order $\delta^{2n}$ to hold exactly, and calculate the
corresponding $n$ coefficients by considering the function $A(T,\delta)$ 
for $n$ different values of $\delta$. In this way we obtain reliable
approximations for the functions $a(T), b(T),...$.
In the present work we employed a $n=3$ point formula 
and used the three smallest dimerizations accessible.
\begin{figure}
\center{\includegraphics[width=\columnwidth]{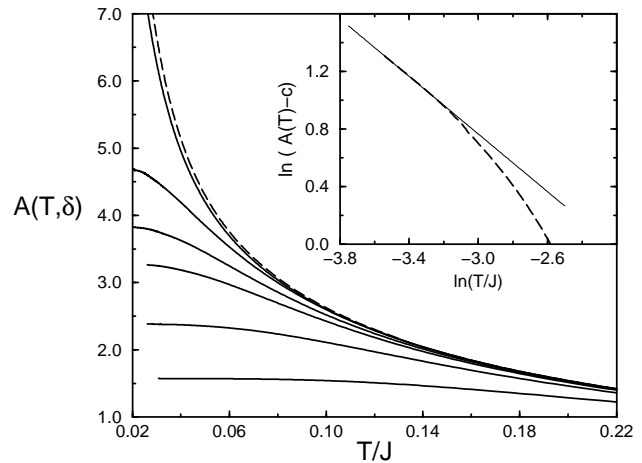}}
\caption[a_von_T0]
   {$A(T,\delta)$ as defined in the text versus temperature for $\alpha =0 $ and 
        $\delta = 0.15, 0.1, 0.05, 0.04, 0.03, 0.01$ (from bottom to top) and the 
        extrapolated $A(T)$ (dashed line). 
        The inset shows a log-log plot of
        $(A(T)+c)$ with $c=-1.14$ (dashed line) together with a 
        line with slope -1 (thin solid line).} 
\label{a_von_T0}
\end{figure}
Here we are particularly interested in $A(T)$ at low temperature
showing a divergence at $T=0$. This singularity is caused by the failure of
(\ref{expans}) at zero temperature. From \cite{Cross79,CF79}
and also from scaling relations of the quantum critical 
Heisenberg chain \cite{Kluemper98} we know for
$\alpha<\alpha_c$ ($\alpha_c \approx 0.2412$ \cite{Okamoto93,Castilla95})
that $f(0,\delta)-f(0,0)\simeq \delta^{4/3}$. For $\alpha>\alpha_c$
the quantum chain shows spontaneous magnetic dimerization as well
as an excitation gap which leads to $f(0,\delta)-f(0,0)\simeq 
\delta$. 

More quantitatively, we first define a temperature variable $t=f(T,0)-f(0,0)$. 
The singular part of the free energy in dependence of $t$ and $\delta$ is denoted by $\tilde{f}(t,\delta) = f(T,\delta) - f(0,0)$. Then, we expect the following scaling relation
\be
\tilde{f}(t,\delta)=\frac{1}{l}\tilde{f}(t l^x,\delta l^y).\label{scal}
\ee
(See e.g. \cite{Fisher82} as a general reference.)
The exponents $x$ and $y$
are to be determined. From (\ref{scal}) we derive in the
usual manner $\tilde{f}(t,\delta)=t^{1/x}\tilde{f}(1,\delta/t^{y/x})$.
Fixing $\delta=0$ directly implies $x=1$ due to the definition of $t$.
Next we use the fact
that $\tilde{f}(0,\delta)\simeq\delta^{\eta}$ which implies $f(1,z)\simeq z^{\eta}$
for large $z$ and furthermore $y=1/\eta$. Lastly we note 
\be
A(T) = \frac{\partial^2}{\partial\delta^2}\tilde{f}(t,0)
=t\frac{1}{t^{2/\eta}}\frac{\partial^2}{\partial\delta^2}\tilde{f}(1,0)
\simeq t^{1-\frac{2}{\eta}}.
\label{scal_low_a}
\ee

Fig.~\ref{a_von_T0} displays $A(T,\delta )$ for $\alpha = 0$ and various 
$\delta$ and $A(T)$ (dashed line) derived from the curves with the lowest 
dimerization. As expected the low-temperature behavior of
the free energy for vanishing dimerization is found to be $t=T^{2}$ with
an extrapolated groundstate energy $f_{0}=-0.443$.  
Here we have $\eta=4/3$, thus we should observe
$A(T)\simeq 1/T$ at sufficiently low $T$. This
is excellently confirmed by our numerics (see inset of
Fig.~\ref{a_von_T0}). As the scaling relations only apply to the
divergent terms we are allowed to subtract regular terms.
For instance an additive constant for
$A(T)$ does not change the scaling, but enlarges the temperature
range where this sets in.

For overcritical frustration $\alpha>\alpha_c$ a similar scaling analysis 
can be performed. The main difference is the absence of quantum criticality
due to the gap $\Delta$ of the elementary
excitations.
\begin{figure}
\center{\includegraphics[width=\columnwidth]{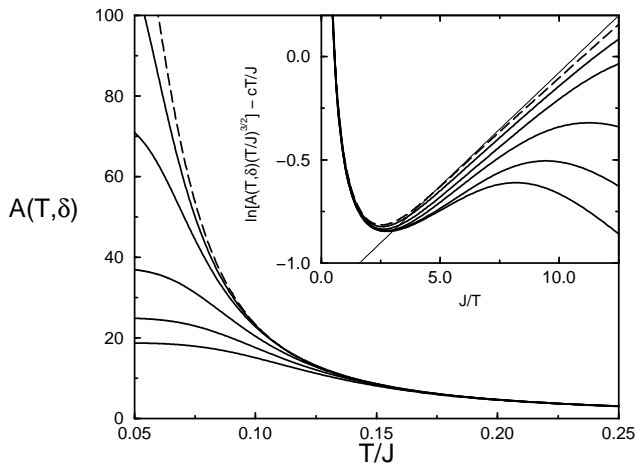}}
\caption[a_von_T]
        {$A(T,\delta)$ versus temperature for $\alpha =0.5$ and 
        $\delta = 0.02, 0.015, 0.01, 0.005, 0.003$ (from bottom to
        top) as well as the 
        extrapolated $A(T)$ (dashed line). The inset shows
        $\ln(A(T,\delta) (T/J)^{3/2}) + c T/J$ with
        $c=0.56$ (solid lines) versus $J/T$, the same transformation of
        the extrapolated $A(T)$
        (dashed line) and a line with slope
        $\Delta_{\mbox{\tiny ST}}/2$ (thin solid line).}
\label{a_von_T}
\end{figure}
Fig.~\ref{a_von_T} shows $A(T,\delta)$ for $\alpha =0.5$ and various $\delta$. 
The coefficient $A(T)$ (dashed line) is deduced as before.
The temperature variable is $t=T^{3/2} \exp(-\Delta/T)$ where
$\Delta=\Delta_{\mbox{\tiny ST}}/2$ is half the singlet-triplet gap
for $\delta=0$.
The origin of this factor $1/2$ can be understood. The elementary 
excitations are spinons each with a gap $\Delta$, however
occurring pairwise. Therefore the spectroscopic data show a gap $2\Delta=
\Delta_{\mbox{\tiny ST}}$. For the usual thermodynamical data like
the free energy the activated behavior shows a characteristic energy 
corresponding to the gap of the individual elementary excitations
whether these occur in arbitrary numbers or restricted to even numbers
does not matter, see e.g. \cite{Kl93}.
(A different situation is realized for 
$\delta>0$ due to the confinement of spinons in pairs.
Here a crossover in the activated behavior sets in at temperatures 
of the order of the binding energy $E_{\mbox{\tiny B}}\leftrightarrow
T_{\mbox{\tiny B}}$. At temperatures above $T_{\mbox{\tiny B}}$ the 
gap in the activated behavior is $\Delta_{\mbox{\tiny ST}}/2$,
below $T_{\mbox{\tiny B}}$ it is $\Delta_{\mbox{\tiny ST}}$.)
For $\delta=0$ the fitted singlet-triplet gap 
$\Delta_{\mbox{\tiny ST}} \approx 0.23(2)$
compares fairly well to the value $0.2338(6)$ \cite{caspe84,Mue-Ha98} for
the Majumdar-Ghosh model. The extrapolated groundstate energy reads
$f_{0} = -0.3749$, also close to the exact result $-\frac{3}{8}$.
Here we have to set $\eta=1$, which implies $A(T)\simeq t^{-1}$.
Indeed we find a linear regime for $\ln[A(T)(T/J)^{3/2}]$ as a function of $1/T$
(see inset of Fig. \ref{a_von_T}), which confirms the prediction of the scaling ansatz (\ref{scal_low_a}). Again, the subtraction of regular terms is admissible. The slight deviations from slope $\Delta_{\mbox{\tiny ST}}/2$ must be apportioned to an inaccuracy in the extrapolated $A(T)$.

Up to now, we have considered static dimerized systems. If $\delta$ is
treated as a thermodynamical degree of freedom, the system favors the 
dimerization which
minimizes the free energy. The spontaneous dimerization $\delta(T)$
as function of
temperature can either be calculated by comparing the free energies
for different dimerizations $\delta$ ($K>0$ fixed) or due
to (\ref{qt}), $\delta(T)$ is implicitly defined by
\be
  \label{impdef}
  A(T, \delta(T)) = \frac{K}{J},
\ee
i.e. the intersection of $A(T,\delta)$ for fixed
$\delta$ with the constant $K/J$. The transition temperature 
$T_{\mbox{\tiny SP}}$ is defined by $\delta(T_{\mbox{\tiny SP}}^-)=0$.
For higher (lower) temperatures the free
energy is increased (lowered) for non-vanishing $\delta$ corresponding
to the uniform (dimerized) phase. For determining
the spontaneous dimerization we use (\ref{impdef}) rather than
the free energy intersection whose accuracy is
expected to be a factor $1/\delta$ worse. 
 
\begin{figure}
\center{\includegraphics[width=\columnwidth]{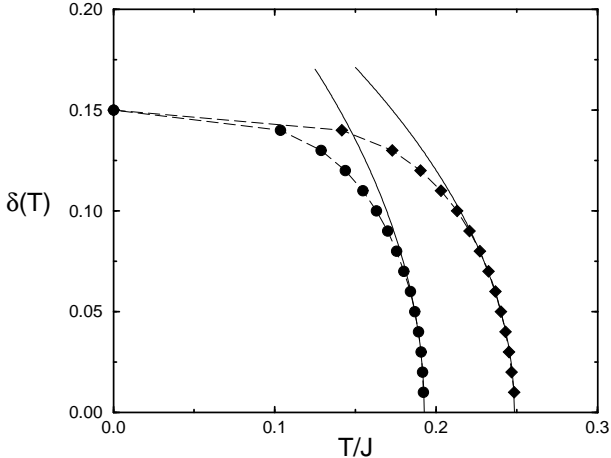}}
\caption[delta_T]{Spontaneous dimerization versus temperature for $\alpha = 0$, $K=1.6J$ 
        (circles) and $\alpha = 0.35$, $K=2.4J$ (diamonds). The lines
        show fitted functions of type $c \cdot (T_{c}-T)^{\beta}$ with $\beta$ fixed at 1/2.}
\label{delta_T}
\end{figure}

In Fig.~\ref{delta_T} the temperature dependence of the spontaneous 
dimerization ($\delta(T)$) is depicted exemplarily for 
 $\alpha = 0, K=1.6J$ and $\alpha = 0.35, K=2.4J$. 
The exponents at the critical temperature $T_{\mbox{\tiny SP}}$ are 
close to $1/2$ which is expected due to the mean-field treatment
of the elastic degrees of freedom. 
The numerical analysis has been performed for a discrete set
of parameters $\delta$ with maximum value $0.15$ which appears to be an upper 
bound for the actual $\delta (T=0)$ of both examples.
The value is in agreement with earlier self-consistent calculations by means 
of the DMRG for zero temperature \cite{Schoen98}.

Moreover, the critical temperature as a function of the elastic constant 
as well as a function of the dimerization $\delta (T=0)$ can 
be deduced from $A(T,\delta)$. The transition 
temperature of \CGO will be discussed in the following section.

\section{Application to \CGO}
Since the discovery of \CGO as the first inorganic spin-Peierls
compound there have been numerous studies of appropriate sets
of microscopic couplings \cite{Castilla95,Riera95,Fab98}. Within the one 
dimensional adiabatic approach commonly used parameter are:
$J \approx 160 K$, $\alpha \approx 0.35$ and 
$\delta (T=0) \approx 1.3 \%$ \cite{Castilla95,Riera95,Fab98}. 
These parameters were derived mainly by fitting the experimental
susceptibility data in the uniform phase well above 
$T_{\mbox{\tiny SP}}$ and from the singlet-triplet gap 
(singlet-singlet gap \cite{Bouzerar97}) 
at zero temperature. Note, that the entropy for these interaction parameters 
also satisfies bounds derived from the specific heat data \cite{Fab98}.

\begin{figure}
\center{\includegraphics[width=\columnwidth]{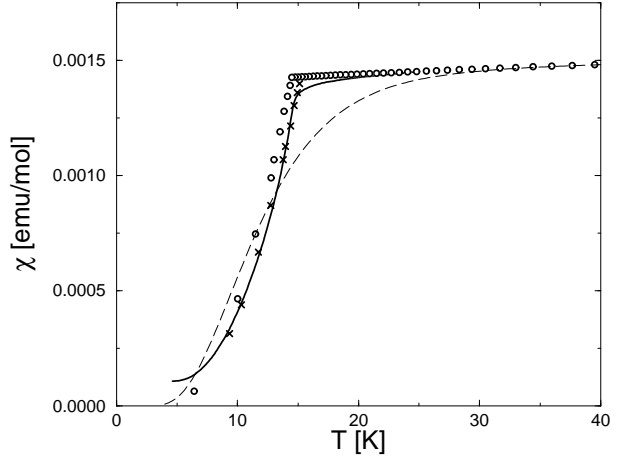}}
\caption[phasediag]
   {Zero field susceptibility of \CGO. Comparison of experimental data
   (solid line) for the $c$-axis ($g$-factor 2.05) with exact
   diagonalization ($L=16$) of the non-dimerized frustrated Heisenberg chain
   (dashed line) and DMRG results using $K=11J$ (circles), $K=10.2J$ (crosses).} 
\label{suscept}
\end{figure}

Next we focus on a brief comparison to experimental data. 
Fig.~\ref{suscept} shows experimental and theoretical susceptibility
data. The agreement for $T>35K$ has already been proven in
\cite{Fab98}. We also find agreement in the region between
$T_{\mbox{\tiny SP}}$ and $35K$.
The deviation close above the phase
transition may be interpreted as a precursor due to fluctuations in the
real system while our theoretical calculations indicate a
sharp transition point.
Using an elastic constant of $K=11J$ the calculated critical temperature is equal to the experimentally determined value $T_{\mbox{\tiny SP}}=14.4K$. Below $T_{\mbox{\tiny SP}}$ the susceptibility is systematically overestimated. With $K=10.2J$, the DMRG
results perfectly coincide with experiment (see Fig.~\ref{suscept}),  
but the transition takes place at about $15.2K$.
Of course, this a natural effect due to the meanfield approximation. A similar overestimation of $T_{\mbox{\tiny SP}}$ is observed by adjusting a square root behavior to the experimentally measured lattice dimerization \cite{Berndpriv98}.

\begin{figure}
\center{\includegraphics[width=\columnwidth]{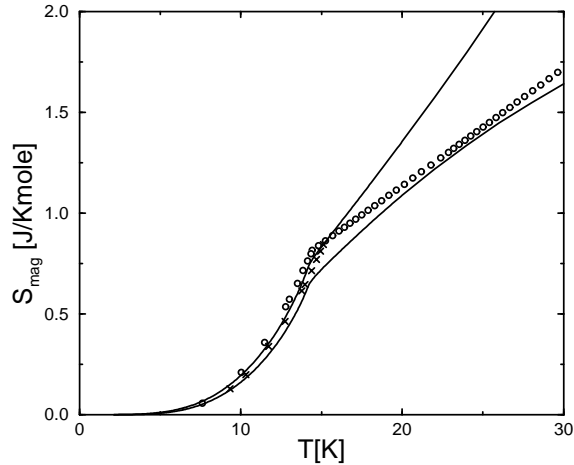}}
\caption[entropy]
   {Entropy for \CGO. Comparison of experimentally determined
   bounds (solid lines) with DMRG results using $K=11J$ (circles) and
   $K=10.2J$ (crosses).} 
\label{entropy}
\end{figure}

In Fig.~\ref{entropy} the entropy $S(T, \delta(T))$ for $\alpha
=0.35$ is depicted. The consistency with the entropy bounds derived 
by an analysis of the specific heat was already mentioned in \cite{Fab98} for
temperatures above $35 K$. The experimentally determined bounds
are also respected for temperatures between $T_{\mbox{\tiny SP}}$ 
and $35K$, below the critical
temperature we observe deviations of about $5\%$ with respect to the
upper bound for $K=11J$. Decreasing the elastic constant to $K=10.2J$
with the consequences stated above the bounds are respected in the
entire temperature range.

Now the low temperature regime is accessible for the first time 
by means of the transfer matrix DMRG. 
To derive the experimentally observed thermodynamics we have to set $K=11J$ $(10.2J)$ which implies a maximum dimerization of 
$\delta(T=0)\approx 2.6\%$ $(2.8\%)$ and $\Delta_{\mbox{\tiny ST}} \approx 40K$ $(42K)$.
The larger dimerization compares well with recent results 
of B\"uchner {\it et al.} \cite{Bernd98}. They conclude a minimum dimerization of
$3\%$ based on measurements of the pressure dependence of the exchange couplings 
together with the structural distortion.
Interestingly, the larger saturation dimerization leads to a singlet-triplet 
gap close to the average gap $\overline{\Delta}_{\mbox{\tiny ST}} \approx 44 K$,
which may be used  within the one dimensional approach regarding the interchain 
interaction ($b$ direction) approximately \cite{Uhrig97}. 

Nevertheless, these elastic constants deviate from earlier 
$T=0$ calculations, where $K=18J$ ($\delta(T=0)\approx 1.4\%$) is found to describe the zero temperature physics correctly \cite{Lorenz98}. 
In particular, the experimentally observed gap and the critical magnetic field at 
the D/I transition are reproduced for $K=18J$.
We have to use a smaller value for $K$ to obtain accordance with 
the experimental transition temperature. The apparent discrepancy of our finite $T$ 
calculations and the results in \cite{Lorenz98} may be explained by 
residual interchain interactions as discussed above.

\section{Summary}
By applying a quantum transfer matrix DMRG algorithm to the
frustrated Heisenberg chain with fixed dimerization we calculated physical 
quantities down to temperatures of less than $0.05$ in units of the coupling 
constant $J$. 

The results compare to analytical predictions based on
scaling arguments with perfect agreement. In particular, we can
clearly distinguish between the quantum critical case ($\alpha <
\alpha_c$) and overcritical frustration ($\alpha > \alpha_c$) which are
characterized by different divergence behavior of the
$\delta^{2}$--coefficient in the free energy. 
Furthermore, the instability towards dimerization can be understood
within this numerical approach. 
It is the first calculation of the spin-Peierls
temperature in the adiabatic model (\ref{Ham})
for realistic sets of parameters $J$, $\alpha$, $K$ 
and the first quantitative investigation of the
dimerized phase at finite temperatures.

Especially with respect to \CGO, taking the parameters $J=160 K$,
$\alpha=0.35$ established by 
high temperature
studies and an elastic constant $K=11J$ $(10.2J)$, a critical temperature of
$T_{\mbox{\tiny SP}}^{\mbox{\tiny theory}}\approx 14.4 K$ $(15.2 K)$ is
found. 
Results from the latter parameter set are in perfect agreement
with the experimental results, e.g. magnetic susceptibility
(Fig.~\ref{suscept}) and entropy (Fig.~\ref{entropy}). The slight
deviation in the value of $T_{\mbox{\tiny SP}}$
can be probably traced back to the static treatment of the lattice
dimerization which overestimates transition temperatures.

In general, our 
results extend the numerical analysis to much lower temperatures,
including the dimerized phase.

\begin{center}
{\bf Acknowledgement}
\end{center}
We like to thank G.S. Uhrig for fruitful discussions, K. Fabricius for providing the 
exact diagonalization results, B. B\"uchner and T. Lorenz for providing the 
experimental data.

The work was supported by the Deutsche Forschungsgemeinschaft through SFB 341.

\end{document}